\documentclass[twocolumn,prl,tightenlines,superscriptaddress,longbibliography]{revtex4-1}
\usepackage{hyperref}
\usepackage{amsmath,amssymb}
\usepackage[utf8]{inputenc} 			
\usepackage[T1]{fontenc}				
\usepackage{graphicx}
\usepackage{float}
\usepackage[usenames,dvipsnames]{xcolor}
\usepackage{tikz}
\usepackage{ulem}

\newcommand{\grad}{\nabla}

\renewcommand{\vec}[1]{\mathbf{#1}}
\newcommand{\Gvec}[1]{\boldsymbol{#1}}

\newcommand*\colvec[3][]{
    \begin{pmatrix}\ifx\relax#1\relax\else#1\\\fi#2\\#3\end{pmatrix}
}

\begin{document}
\title{Self-Organized Critical Coexistence Phase in Repulsive Active Particles}

\author{Xia-qing Shi}
\affiliation{Center for Soft Condensed Matter Physics and Interdisciplinary Research, Soochow University, Suzhou 215006, China}
\affiliation{Service de Physique de l'Etat Condens\'e, CEA, CNRS Universit\'e Paris-Saclay, CEA-Saclay, 91191 Gif-sur-Yvette, France}

\author{Giordano Fausti}
\affiliation{Service de Physique de l'Etat Condens\'e, CEA, CNRS Universit\'e Paris-Saclay, CEA-Saclay, 91191 Gif-sur-Yvette, France}

\author{Hugues Chat\'{e}}
\affiliation{Service de Physique de l'Etat Condens\'e, CEA, CNRS Universit\'e Paris-Saclay, CEA-Saclay, 91191 Gif-sur-Yvette, France}
\affiliation{Computational Science Research Center, Beijing 100094, China}
\affiliation{Sorbonne Universit\'e, CNRS, Laboratoire de Physique Th\'eorique de la Mati\`ere Condens\'ee, 75005 Paris, France}

\author{Cesare Nardini}
\affiliation{Service de Physique de l'Etat Condens\'e, CEA, CNRS Universit\'e Paris-Saclay, CEA-Saclay, 91191 Gif-sur-Yvette, France}

\author{Alexandre Solon}
\affiliation{Sorbonne Universit\'e, CNRS, Laboratoire de Physique Th\'eorique de la Mati\`ere Condens\'ee, 75005 Paris, France}

\date{\today}
\begin{abstract}
  We revisit motility-induced phase separation in two models of active
  particles interacting by pairwise repulsion.  We show that the
  resulting dense phase contains gas bubbles distributed algebraically
  up to a typically large cutoff scale. At large enough system size
  and/or global density, all the gas may be contained inside the
  bubbles, at which point the system is microphase-separated with a
  finite cut-off bubble scale. We observe that the ordering is
  anomalous, with different dynamics for the coarsening of the dense
  phase and of the gas bubbles. This phenomenology is reproduced by a
  ``reduced bubble model'' that implements the basic idea of reverse
  Ostwald ripening put forward in Tjhung {\it et al.}  [Phys. Rev. X
  {\bf 8}, 031080 (2018)].
\end{abstract}
\maketitle

Self-propelled particles interacting solely with steric repulsion are
well known to be able to spontaneously separate into a macroscopic
dense cluster and a residual gas, in spite of the absence of explicit
attraction forces. This motility-induced phase separation
(MIPS)~\cite{cates_motility-induced_2015} of active particles has
become a cornerstone of the physics of dry active matter (in which the
fluid surrounding particles is neglected).  As such, it has driven
many theoretical
works~\cite{tailleur_statistical_2008,stenhammar_continuum_2013,wittkowski_scalar_2014,solon_pressure_2015,solon_generalized_2018,tjhung_cluster_2018}
as well as countless numerical studies (see
e.g. \cite{fily_athermal_2012,redner_structure_2013,speck_effective_2014,stenhammar_phase_2014,digregorio_full_2018,klamser_thermodynamic_2018,mandal_motility_2019,caprini_spontaneous_2020}
to name a few prominent ones).  The motility reduction resulting from
persistent collisions, which leads to MIPS, is a generic ingredient
encountered both in living and synthetic active
matter~\cite{liu_self-driven_2019,buttinoni_dynamical_2013,geyer_freezing_2019,vanderlinden_interrupted_2019}.

Despite its purely non-equilibrium origin, MIPS was initially
described as a conventional phase separation between two homogeneous
macroscopic phases. It was first predicted in models of quorum-sensing
particles~\cite{tailleur_statistical_2008} where particle speed
decreases with the local density, without two-body interactions. In
this case, equilibrium-like thermodynamics can be constructed to
account quantitatively for phase
coexistence~\cite{solon_generalized_2018-1}. For systems of repulsive
disks, attempts were made to model the speed reduction due to
collisions by a quorum-sensing
interaction~\cite{fily_athermal_2012,stenhammar_continuum_2013,speck_effective_2014},
but the results are not satisfactory, due to fundamental differences
between the two types of
interactions~\cite{solon_pressure_2015-2,solon_pressure_2015}.

There is indeed mounting evidence that more complex physics is at play
in systems of repulsive disks. In particular, the surface tension
between the dense phase and the gas, defined via the Laplace law, has
been measured to be
negative~\cite{bialke_negative_2015,solon_generalized_2018-1,patch_curvature-dependent_2018},
triggering a spate of controversy\footnote{Two recent papers either
  contest~\cite{hermann_non-negative_2019} or regard as physically
  irrelevant~\cite{omar_microscopic_2020} that the surface tension is
  measured to be negative in numerical simulations.}.  This was
rationalized at field theoretical level by including terms that break
detailed balance in the classical theory for equilibrium liquid-gas
phase separation.  In this active Model B+ (AMB+), surface tension can
become negative for some parameter values, in which case Ostwald
ripening is reversed for vapor bubbles while still remaining normal
for liquid droplets. This means that small vapor bubbles, contrary to
the standard scenario, grow at the expense of larger ones due to a
diffusion flux. When this happens, simulations of AMB+ lead to either
a bubbly fluid interpreted as microphase separation, or to the
coexistence of a dense phase populated of bubbles with an outer
gas~\cite{tjhung_cluster_2018}. There is in fact incidental evidence
for such a bubbly liquid at particle
level~\cite{redner_structure_2013,bialke_negative_2015,stenhammar_phase_2014,solon_generalized_2018-1,digregorio_full_2018},
but it has not yet been studied {\it per se}. Very recently, Caporusso
et al~\cite{caporusso2020micro} have shown more clearly that in
systems with hard-core interactions, the dense phase is made of
hexatic subdomains and interstitial gas regions.

In this Letter we show, within two standard particle models displaying
MIPS, that not only the dense phase is endowed with bubbles, but also
that these are distributed algebraically up to some cutoff scale that
we observe to grow with system size. Finite-size scaling based on this
observation suggests that, as system size increases, more and more of
the gas is contained in bubbles. At large densities, we are able to
observe the vanishing of the macroscopic gas reservoir, and the system
is then microphase-separated with bubbles of all sizes up to a maximal
bubble size that depends on the average density. Moreover, the
coarsening of bubbles is anomalous with the typical length scale
growing as $t^{0.22}$. We elucidate the basic mechanisms at play, and
show, within a reduced model implementing reversed Ostwald ripening
for gas bubbles, that they indeed lead to a self-organized critical
dynamics.

{\it Self-organized critical phase coexistence.}  We first consider
the paradigmatic active brownian particles (ABPs) introduced in \cite{fily_athermal_2012}. 
Self-propelled by a force of constant magnitude $F_0$ along its internal polarity
$\vec u_i=(\cos\theta_i,\sin\theta_i)$, particle $i$ evolves according
to the overdamped Langevin equations governing its position
${\bf r}_i$ and polar angle $\theta_i$:
\begin{equation}
  \dot{\vec r}_i = {\Gvec \mu}_i (F_0 {\vec u}_i +  {\vec F}_i) +  {\Gvec \eta}_i\; ;\qquad \dot{\theta}_i = \eta_i
  \label{eq:Langevin-ABP}
\end{equation}
where ${\vec F}_i=-\sum_{j\neq i} \grad V(\vec r_i-\vec r_j)$ is the
force exerted on particle $i$ by the other particles. We choose the
pair potential to be a short-range harmonic repulsion
$V(r)=\frac{k}{2}(\sigma-r)^2$ if $r<\sigma$ and $0$ otherwise with
$k$ the repulsive strength and $\sigma=1$ the interaction radius.
In contrast to previous studies, we allow the mobility tensor
${\Gvec \mu}_i$ and the translational noise ${\Gvec \eta}_i$ to be
anisotropic, as expected generically for active particles:
${\Gvec \mu}_i = \mu_\|{\vec u}_i{\vec u}_i+\mu_\perp({\rm I}-{\vec
  u}_i{\vec u}_i)$,
$\Gvec{\eta}_i = \sqrt{2\tau\mu_\|} \xi^\|_i {\vec u}_i
+\sqrt{2\tau\mu_\perp} \xi^\perp_i ({\vec u}_i \times {\vec z})$, and
$\eta_i= \sqrt{2\tau\mu_\theta} \xi^\theta_i$ with $\tau$ a parameter
controlling the noise strength, ${\vec z}$ the unit vector
perpendicular to the plane of motion, and the $\xi_i$'s Gaussian white
noises with unit variance.

\begin{figure}
  \centering
  \includegraphics[width=1\linewidth]{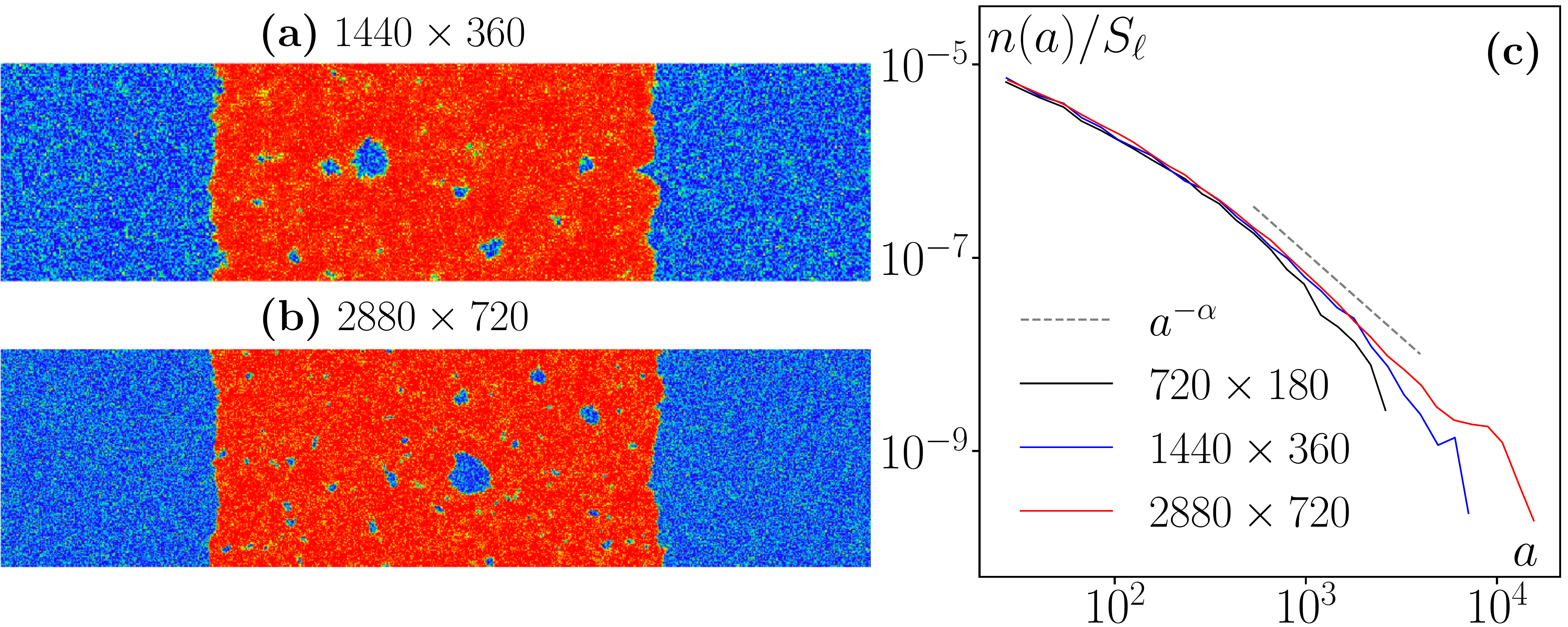}
  \caption{Active Brownian particles.
  Typical snapshots in steady state at system size $S=1440\times 360$ (a)
    and $S=2880\times 720$ (b). (Colors represent the packing fraction calculated over $2\times 2$ boxes, from 0 (dark blue) to 1 (red).)
    (c): Rescaled distribution of bubble area $n(a)/S_\ell$ at various system sizes (indicated in legends).  
    Parameters: $\rho_0=\tfrac{N\pi}{4S}=0.6$, $\tau=0.01$, $\mu_\perp=1/4$, $\mu_\|=1$,
    $\mu_\theta=3/2$, $F_0=1$ and $k=20$. Typical averaging time is $4\times10^6$ after a transient of $10^6$.
    }     
  \label{fig1}
\end{figure}

All simulations below are of large two-dimensional domains with
periodic boundary conditions. Numerical details are given in
\cite{SUPP}.  At phase coexistence, we observe, inside the macroscopic
dense domain, persistent bubbles with a range of sizes, surrounded by
a liquid (see movie in~\cite{SUPP}). Bubbles are more prominent when
the mobility is anisotropic~\footnote{This may explain why previous
  works, all performed with isotropic mobility, did not pay much
  attention to bubbles that remained rare and small at the sizes
  studied.}.  Typical snapshots for $\mu_\|=4\mu_\perp$ and
$\mu_\theta=6\mu_\perp$ are shown in Fig.~\ref{fig1}(a,b) for systems
only differing by their size. Clearly, doubling system size increases
the size of the bubbles.  Fig.~\ref{fig1}(c) shows, at different
system sizes, $n(a)$, the average number of bubbles of area $a$,
normalized by $S_{\ell}$, the total area of the liquid in which
bubbles live. The distributions collapse on an increasing range, span
several orders of magnitude, and decay approximately as a power law
$n(a)\sim a^{-\alpha}$ with $\alpha\approx 1.75$ terminated by a
cutoff that {\it increases} with system size.

The ABP simulations reported above only show a rather short scaling
range.  Numerically, the main limitation is not so much system size
than the huge times needed to obtain clean averages~\footnote{This is illustrated in Fig.~\ref{fig3}(a) for the other model studied here.}.  We thus implemented an active lattice gas~\cite{thompson_lattice_2011,soto2014self,sepulveda_wetting_2017,whitelam_phase_2018,partridge_critical_2019}:
on an hexagonal lattice~\cite{partridge_critical_2019}, particles
carrying an internal polarity pointing to one of the 6 lattice
directions attempt to perform one of 3 moves (see \cite{SUPP} for details).
(i) With a rate $r_P$ they perform a
`self-propelled' jump to the nearest site along their internal
polarity direction.  (ii) They undergo spatial diffusion to any neighboring
site at rate $r_D$, and (iii) rotational diffusion (changing their polarity
to one of its two neighboring orientations) at rate $r_R$.  For
optimal efficiency, we impose strict exclusion and parallel updating.

\begin{figure*}
  \centering
 \includegraphics[width=\linewidth]{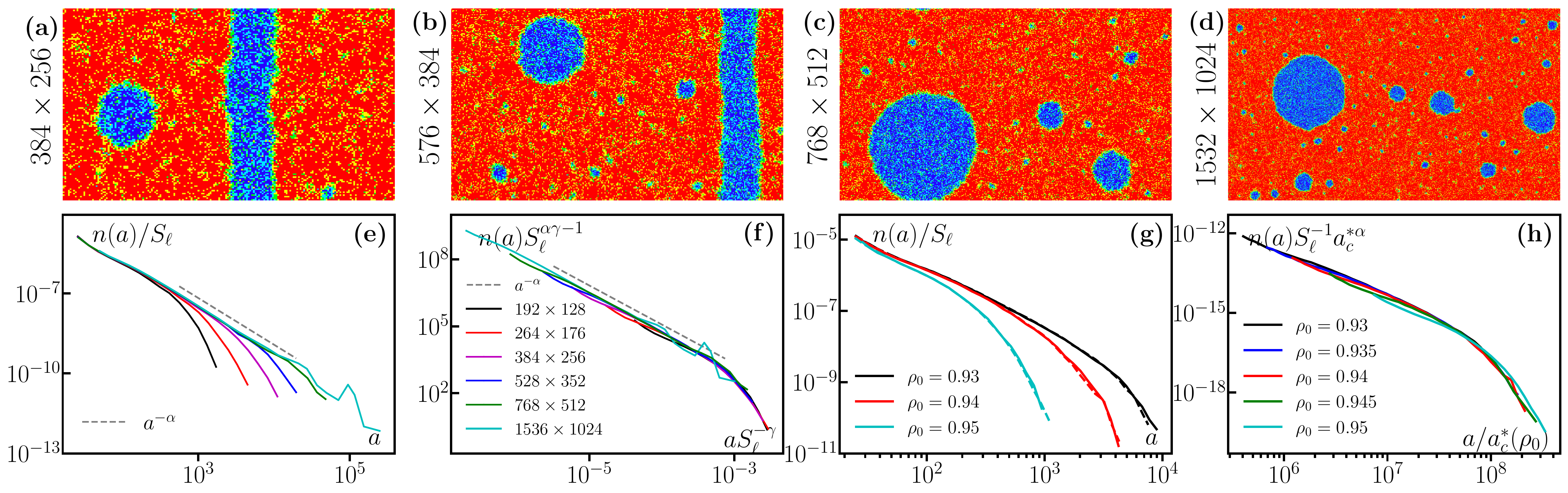}
  \caption{Active lattice gas. 
  (a-d) Snapshots in the steady state at $\rho_0=0.8$ in systems of different sizes from 
  $S=384\times 256$ to $S=1532\times 1024$ (colors as in Fig.~\ref{fig1}(a,b)). 
  For the two biggest sizes (c,d), the system is in the microphase-separated regime.
  (e-h) bubble area distribution in the SOC scaling regime (e,f: $\rho_0=0.6$) 
  and in the microphase-separated regime (g,h). 
  (e) $n(a)/S_\ell$ at various system sizes (indicated by the legends in (f)). 
  Typical averaging time is $10^{10}$ timesteps after discarding a transient of $10^8$. 
  (f): same as (e), but as function of $a/S_\ell^\gamma$ with $\gamma=1.40$. 
  (g): $n(a)/S_\ell$ at different $\rho_0$ values for $S=768\times 1024$ (dashed
    lines) and $S=1536\times 2048$ (solid lines).
 (h): same as (g), but rescaled according to Eq.~(\ref{eq:cutoff-microp}).
  The grey dashed lines have slope $-\alpha=-1.75$.
  }
  \label{fig2}
\end{figure*}

In the following, we use $r_P=1$, $r_D=2$ and $r_R=0.032$, typical
values leading to phase separation (a study of the phase diagram will
be presented elsewhere).  Persistent bubbles are clearly visible
(Fig.~\ref{fig2}(a-d)).  The bubble area distribution in the globally
phase-separated regime is similar to that observed for ABPs but with a
much larger scaling region (Fig.~\ref{fig2}(e)):
$n(a)\propto a^{-\alpha}$ with $\alpha=1.75(5)$. This region extends
to a cut-off size $a_c$ that grows with the total liquid area as
$a_c\propto S_\ell^{\gamma}$ with $\gamma=1.40(5)$. The distributions
can thus be collapsed on a master curve using these two exponents
(Fig.~\ref{fig2}(f)).

In both models presented, we find that whenever the system is globally
phase separated, the dense phase contains bubbles.  This phase bears
the hallmarks of self-organized criticality (SOC) (for recent
overviews, see \cite{Pruessner_SOC_2012,buenda2020feedback}).  Small
bubbles are nucleated inside the liquid, diffuse and grow by merging
with other bubbles. This process get slower and slower with increasing
bubble size.  Bubbles are eventually expelled into the reservoir of
outside gas upon touching the boundary, in sudden, avalanche-like
events, providing separation of timescales (see movie in \cite{SUPP}).
As in typical SOC systems, avalanches occur at all accessible scales.

\begin{figure}
  \centering
  \includegraphics[width=\linewidth]{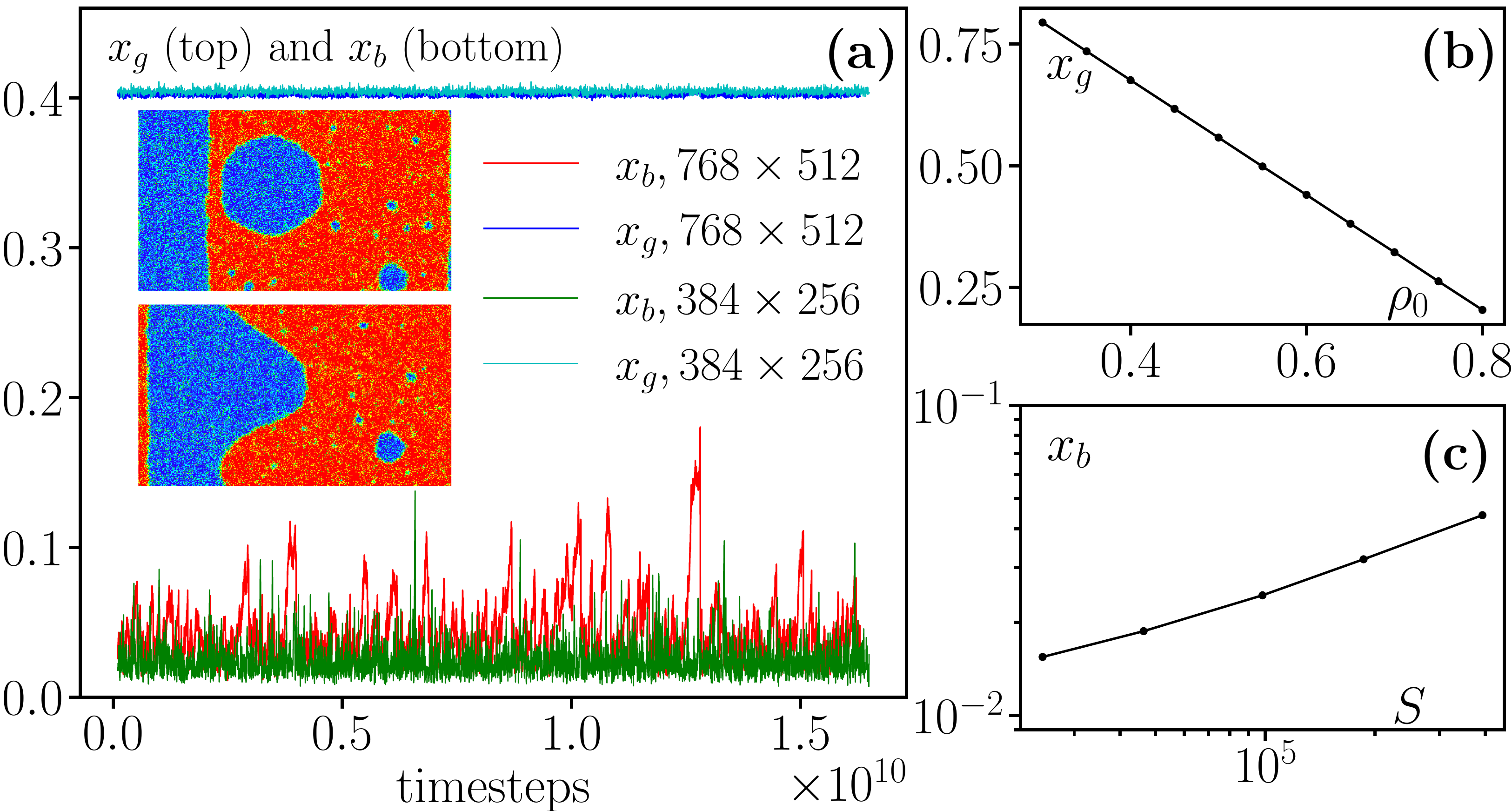}
  \caption{Active lattice gas at $\rho_0=0.63$.
  (a): timeseries of the area fraction occupied by bubbles
    $x_b$ (bottom) and by total gas $x_g$ (top) for
    system sizes $S=384\times 256$ and $S=768\times 512$; 
    insets: two snapshots of the system taken right before and at the sharpest peak in the $x_b$ timeseries at 
    $S=768\times 512$ (red curve, around $t=1.3\times 10^{10}$).
    (b): Linear variation of $x_g$ with $\rho_0$ (lever rule) computed for $S=384\times256$.
    (c): Bubble fraction $x_b$ v system size $S$.
    }
  \label{fig3}
\end{figure}

The SOC-like mechanisms leading to an algebraic distribution of
bubbles do {\it not} invalidate the global picture of a phase
separation between gas and liquid with fixed densities $\rho_g$ and
$\rho_\ell$ independent of system size up to small finite-size
corrections. However, the gaseous part of the system is formed here of
the outside gas reservoir {\it and} of the bubbles. With this definition,
the gas fraction $x_g$ fluctuates very little and is independent of
system size to a good approximation (Fig.~\ref{fig3}(a)).  Moreover,
$x_g$ varies linearly with the average density $\rho_0$
(Fig.~\ref{fig3}(b)) so that the lever rule still applies: For
$\rho_g<\rho_0<\rho_\ell$, the average density sets the fraction of
liquid and gas in the system
$x_g=(\rho_\ell-\rho_0)/(\rho_\ell-\rho_g)$ and
$x_\ell=1-x_g$.  The fluctuations of $x_b$, the fraction of
the system occupied by bubbles, in contrast to the gentle ones of
$x_g$, are large, intermittent, and increase with system
size. (Note the huge timescales over which fluctuations of $x_b$
occur even at the modest sizes shown in Fig.~\ref{fig3}(a).)  Their
stronger and stronger peaks reflect the larger and larger avalanches
(expulsion of bubbles) (Fig.~\ref{fig3}(a), insets).

{\it Microphase-separated bubbly liquid.} The lever rule immediately
tells us that the SOC scaling evidenced above cannot continue
asymptotically when system size $S\to\infty$. Indeed, 
the bubble area fraction grows with system size:
\begin{equation}
  \label{eq:phib}
 x_b\equiv \frac{1}{S}\int_0^\infty \!\! a\, n(a) da\approx \frac{S_\ell}{S}\int_0^{a_c} \!\!  a^{1-\alpha} da\propto x_\ell S_\ell^{\gamma(2-\alpha)},
\end{equation}
where we used the scalings of $n(a)$ and $a_c(S_\ell)$.  Our numerical
data confirm this (Fig.~\ref{fig3}(c)).  Surely,
Eq.~(\ref{eq:phib}) ceases to be possible once all the gas is
contained in the bubbles, $x_b=x_g$, which happens at a typical
crossover size
$S^*\propto (x_g/x_\ell)^{1/[\gamma(2-\alpha)]}/x_\ell$. The cutoff on
bubble size then reads
\begin{equation}
  \label{eq:cutoff-microp}
  a_c^*\equiv a_c(S^*)\propto (x_g/x_\ell)^{1/(2-\alpha)}.
\end{equation}
Equation~\eqref{eq:cutoff-microp} implies that $S^*$ and $a_c^*$
depend on the average density $\rho_0$ through $x_g$ and $x_\ell$ and
that they diverge near the gas binodal $\rho_0\to\rho_g$.  On the
other hand they get smaller when approaching the liquid binodal.
Beyond $S^*$ the system settles in a micro-phase separated state, a
homogeneous liquid with bubbles of all sizes up to $a_c^*$.  SOC
scaling then breaks down, and $n(a)$ becomes independent of system
size.

Using our lattice gas model at high enough $\rho_0$, we are able to
reach system sizes where all the gas is contained in bubbles, and the
system settles in the micro-phase separated state
(Fig.~\ref{fig2}). Our data are in agreement with our scaling
arguments: $n(a)$ is then independent of system size and is cut off at
some scale that depends only on the average density
(Fig.~\ref{fig2}(g)). Plotting $n(a/a_c^*)$ collapses the
distributions for different $\rho_0$, confirming the validity of
Eq.~(\ref{eq:cutoff-microp}) (Fig.~\ref{fig2}(h)), at least close to
the liquid binodal.

{\it Reduced bubble model.} The AMB+ field theory of
Ref.~\cite{tjhung_cluster_2018} suggests that bubbles exist because of
reverse Ostwald ripening, which causes large bubbles to shrink at the
advantage of small ones, thus competing with coalescence. To test
whether these ingredients are sufficient to reproduce the
phenomenology described above, we implemented them in a reduced model
whose degrees of freedom are the positions and radii of bubbles that
we assume to be perfectly circular~\footnote{For a similar approach in
  a different context, see \cite{ranft_aggregation_2017}.}.

The bubble-particles evolve in continuous time in a continuous domain.
New bubbles with radius $r_0=1$ are nucleated in the liquid at a small
rate $k_n$ per unit area. In line with the reverse Ostwald scenario,
the new bubbles are nucleated at the expense of the larger ones: all
other bubbles shrink by an amount $\kappa r(1-r_0/r)$ (where $r$ is
their current radius), with $\kappa$ chosen such that the total area
of gas is conserved. (Note that this neglects spatial effects: In
principle, bubbles would equilibrate in priority with neighboring
ones.)  Bubbles chosen randomly among the current $n(t)$ existing ones
diffuse with a coefficient $D$, that for simplicity we assume
constant.  If the move brings the bubble into contact with another,
they merge into a single one located at their ``center of mass",
conserving total area.  To have the same geometry as in globally
phase-separated microscopic models, we also add the possibility to
have a gas reservoir outside two parallel interfaces that move along
the dynamics to insure that $x_\ell$ remains constant.

\begin{figure}
  \centering
  \includegraphics[width=\linewidth]{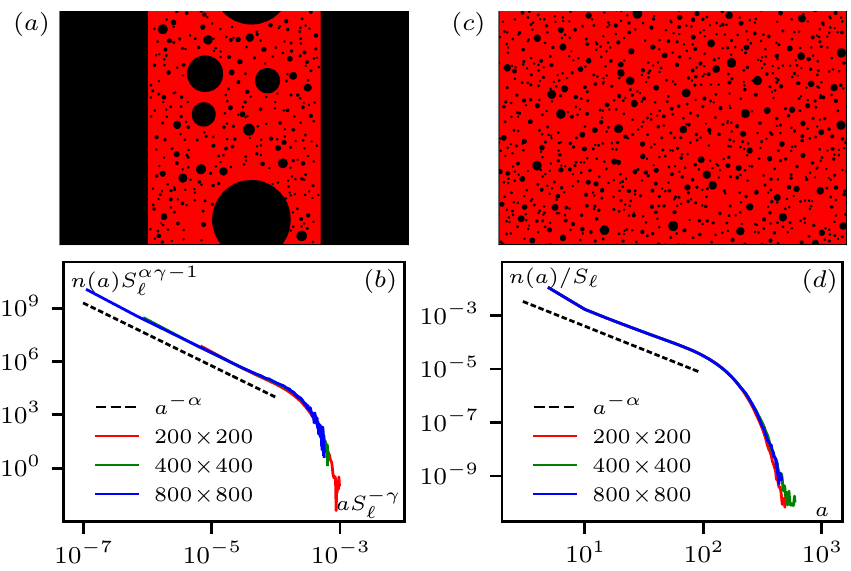}
  \caption{Reduced bubble model. 
  (a,c): Typical snapshots at
    $x_g=0.7$ in the SOC coexistence regime (a) and at
    $x_g=0.05$ in the microphase regime (c). System size
    $S=600\times 400$. The liquid and gas are represented in red and
    black respectively. 
    (b,d)  Bubble area distribution in
    the two regimes, rescaled using $\alpha=1.77$, $\gamma=1.48$. For
    all panels $D=1$, $k_n=10^{-4}$.}
  \label{fig:Bmodel}
\end{figure}

While a complete presentation of the behavior of this reduced model
and some variants will be reported elsewhere, here we show that it
typically yields a phenomenology remarkably similar to that described
above.  Fixing $D=1$ and $k_n=10^{-4}$, we vary $x_g$ the total gas
fraction and the system size $S$.  At high $x_g$ or small system size
$S$, we observe the SOC coexistence between a bubbly liquid and a gas
reservoir (Fig.~\ref{fig:Bmodel}(a,b) and supplementary
movie~\cite{SUPP}).  The bubble size distribution scales as in
Fig.~\ref{fig2}(f) with exponent values close to those of the lattice
gas ($\alpha=1.77(2)$ and $\gamma=1.48(5)$), but preliminary results
(not shown) suggest that they are not universal.  Decreasing $x_g$ or
increasing $S$, the gas reservoir becomes smaller and smaller until it
disappears, at which point we have a microphase separated regime with
$n(a)$ independent of system size (Fig.~\ref{fig:Bmodel}(c,d)).
Despite its simplicity, our bubble model thus captures the essential
phenomenology described here.

\begin{figure}[t!]
  \centering
  \includegraphics[width=\linewidth]{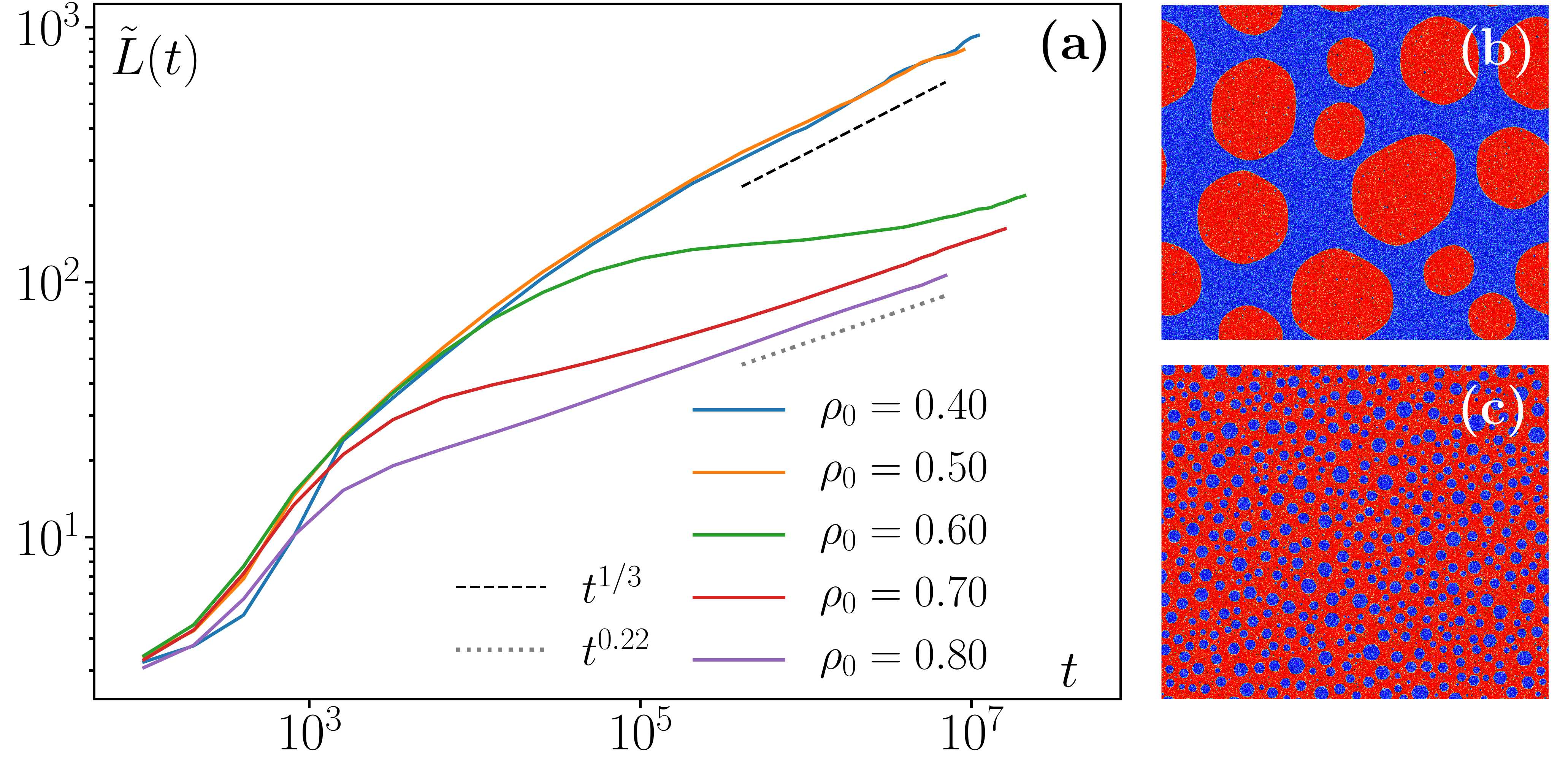}
  \caption{Coarsening process in the active lattice gas
    ($S=8192 \times 6144$).  (a) Timeseries of the typical lengthscale
    $\tilde{L}$ starting from random initial conditions at different
    $\rho_0$ values.  (b,c) Snapshots of the system taken at $t=10^7$
    for $\rho_0=0.5$ (b) and $0.7$ (c). At all times shown here,
    $\tilde{L}\ll \sqrt{S}$.  }
  \label{fig:coarsening}
\end{figure}

{\it Coarsening process.} We finally study the growth of order
following random initial conditions, considering the characteristic
length extracted from the structure factor \cite{SUPP}. We only
present results for our active lattice gas
(Fig.~\ref{fig:coarsening}), but similar ones, albeit of lesser
quality, were obtained for ABPs. When the liquid is the majority
phase, after an initial transient, the coarsening is dominated by
vapor bubbles and follows an anomalous $t^{0.22}$ law. We currently
lack an analytical explanation of such law.  Instead, when the liquid
is the minority phase, it is dominated by liquid droplets and
coarsening is normal $t^{1/3}$ as expected both from Ostwald ripening
and coalescence in models with conserved order
parameter~\cite{bray2002theory}. In fact, within the liquid droplets
we expect the bubbles to coarsen as well with the anomalous law
above. However, being slower than the liquid coarsening, it is not
surprising that this is not visible in our data.

{\it Conclusion.}  We have shown, using two very different models of
active particles interacting strictly by pairwise repulsion, that the
dense phase resulting from MIPS is critical, containing bubbles of gas
distributed algebraically up to some cutoff scale. We observe at high
density that, as long as an outer gas phase is present, this cutoff
increases as a power of system size. At large enough system size
and/or global density, the gas reservoir may disappear and the cutoff
scale becomes independent of system size. This asymptotic regime is
thus microphase separated.  A ``reduced bubble model'' captures this
essential phenomenology within a minimal framework that implements the
basic idea of reverse Ostwald ripening put forward in
Ref.~\cite{tjhung_cluster_2018}.

In the models presented here, the asymptotic cutoff scale $a_c^*$
grows very fast with $\rho_0$.  Numerically, we are only able to
access the asymptotic regime at rather high density (Fig.~\ref{fig2}).
What happens asymptotically at low densities thus remains unknown, but
extrapolating the scaling laws uncovered here lead us to speculate
that our scenario remains valid in the whole phase coexistence region
$\rho_g<\rho_0<\rho_\ell$.

This Letter leaves several important open questions. In particular,
whether the scenario described here is observed whenever pairwise
repulsion is present, and whether the critical exponents are universal
could be addressed numerically by considering other models showing
MIPS, the AMB+ field theory, and reduced bubble models with different
parameters.
In this context, the very recent work of Caporusso et
al~\cite{caporusso2020micro}, where a very hard potential between ABPs
leads to crystalline clusters that aggregate to form a dense phase
with interstitial gas, might be understood within our scenario.
Finally, in regards to the current controversy about the nature of the
critical point of
MIPS~\cite{siebert_critical_2018,partridge_critical_2019}, our results
make it unlikely that it belongs to the Ising universality class, but
this hard problem remains thus unsettled.

\begin{acknowledgements}
  We thank M.E. Cates, A. Patelli and J. Tailleur for interesting
  discussions related to this work. This work is supported by the
  National Natural Science Foundation of China (Grant No. 11635002 to
  X.-q.S. and H.C., Grants No. 11922506 and No. 11674236 to
  X.-q.S.). C.N. acknowledges the support of an Aide Investissements
  d’Avenir du LabEx PALM (ANR-10-LABX-0039-PALM).  G.F. was supported
  by the CEA NUMERICS program, which has received funding from the
  European Union's Horizon 2020 research and innovation program under
  the Marie Sklodowska-Curie grant agreement No 800945.
\end{acknowledgements}

\bibliography{bubble-refs.bib}

\begin{thebibliography}{41}%
\makeatletter
\providecommand \@ifxundefined [1]{%
 \@ifx{#1\undefined}
}%
\providecommand \@ifnum [1]{%
 \ifnum #1\expandafter \@firstoftwo
 \else \expandafter \@secondoftwo
 \fi
}%
\providecommand \@ifx [1]{%
 \ifx #1\expandafter \@firstoftwo
 \else \expandafter \@secondoftwo
 \fi
}%
\providecommand \natexlab [1]{#1}%
\providecommand \enquote  [1]{``#1''}%
\providecommand \bibnamefont  [1]{#1}%
\providecommand \bibfnamefont [1]{#1}%
\providecommand \citenamefont [1]{#1}%
\providecommand \href@noop [0]{\@secondoftwo}%
\providecommand \href [0]{\begingroup \@sanitize@url \@href}%
\providecommand \@href[1]{\@@startlink{#1}\@@href}%
\providecommand \@@href[1]{\endgroup#1\@@endlink}%
\providecommand \@sanitize@url [0]{\catcode `\\12\catcode `\$12\catcode
  `\&12\catcode `\#12\catcode `\^12\catcode `\_12\catcode `\%12\relax}%
\providecommand \@@startlink[1]{}%
\providecommand \@@endlink[0]{}%
\providecommand \url  [0]{\begingroup\@sanitize@url \@url }%
\providecommand \@url [1]{\endgroup\@href {#1}{\urlprefix }}%
\providecommand \urlprefix  [0]{URL }%
\providecommand \Eprint [0]{\href }%
\providecommand \doibase [0]{http://dx.doi.org/}%
\providecommand \selectlanguage [0]{\@gobble}%
\providecommand \bibinfo  [0]{\@secondoftwo}%
\providecommand \bibfield  [0]{\@secondoftwo}%
\providecommand \translation [1]{[#1]}%
\providecommand \BibitemOpen [0]{}%
\providecommand \bibitemStop [0]{}%
\providecommand \bibitemNoStop [0]{.\EOS\space}%
\providecommand \EOS [0]{\spacefactor3000\relax}%
\providecommand \BibitemShut  [1]{\csname bibitem#1\endcsname}%
\let\auto@bib@innerbib\@empty
\bibitem [{\citenamefont {Cates}\ and\ \citenamefont
  {Tailleur}(2015)}]{cates_motility-induced_2015}%
  \BibitemOpen
  \bibfield  {author} {\bibinfo {author} {\bibfnamefont {M.~E.}\ \bibnamefont
  {Cates}}\ and\ \bibinfo {author} {\bibfnamefont {J.}~\bibnamefont
  {Tailleur}},\ }\bibfield  {title} {\enquote {\bibinfo {title}
  {Motility-{Induced} {Phase} {Separation}},}\ }\href {\doibase
  10.1146/annurev-conmatphys-031214-014710} {\bibfield  {journal} {\bibinfo
  {journal} {Annu. Rev. Condens. Matter Phys}\ }\textbf {\bibinfo {volume}
  {6}},\ \bibinfo {pages} {219--244} (\bibinfo {year} {2015})}\BibitemShut
  {NoStop}%
\bibitem [{\citenamefont {Tailleur}\ and\ \citenamefont
  {Cates}(2008)}]{tailleur_statistical_2008}%
  \BibitemOpen
  \bibfield  {author} {\bibinfo {author} {\bibfnamefont {J.}~\bibnamefont
  {Tailleur}}\ and\ \bibinfo {author} {\bibfnamefont {M.~E.}\ \bibnamefont
  {Cates}},\ }\bibfield  {title} {\enquote {\bibinfo {title} {Statistical
  mechanics of interacting run-and-tumble bacteria},}\ }\href {\doibase
  10.1103/physrevlett.100.218103} {\bibfield  {journal} {\bibinfo  {journal}
  {Phys. Rev. Lett.}\ }\textbf {\bibinfo {volume} {100}},\ \bibinfo {pages}
  {218103} (\bibinfo {year} {2008})}\BibitemShut {NoStop}%
\bibitem [{\citenamefont {Stenhammar}\ \emph {et~al.}(2013)\citenamefont
  {Stenhammar}, \citenamefont {Tiribocchi}, \citenamefont {Allen},
  \citenamefont {Marenduzzo},\ and\ \citenamefont
  {Cates}}]{stenhammar_continuum_2013}%
  \BibitemOpen
  \bibfield  {author} {\bibinfo {author} {\bibfnamefont {J.}~\bibnamefont
  {Stenhammar}}, \bibinfo {author} {\bibfnamefont {A.}~\bibnamefont
  {Tiribocchi}}, \bibinfo {author} {\bibfnamefont {R.~J.}\ \bibnamefont
  {Allen}}, \bibinfo {author} {\bibfnamefont {D.}~\bibnamefont {Marenduzzo}}, \
  and\ \bibinfo {author} {\bibfnamefont {M.~E.}\ \bibnamefont {Cates}},\
  }\bibfield  {title} {\enquote {\bibinfo {title} {Continuum theory of phase
  separation kinetics for active brownian particles},}\ }\href {\doibase
  10.1103/physrevlett.111.145702} {\bibfield  {journal} {\bibinfo  {journal}
  {Phys. Rev. Lett.}\ }\textbf {\bibinfo {volume} {111}},\ \bibinfo {pages}
  {145702} (\bibinfo {year} {2013})}\BibitemShut {NoStop}%
\bibitem [{\citenamefont {Wittkowski}\ \emph {et~al.}(2014)\citenamefont
  {Wittkowski}, \citenamefont {Tiribocchi}, \citenamefont {Stenhammar},
  \citenamefont {Allen}, \citenamefont {Marenduzzo},\ and\ \citenamefont
  {Cates}}]{wittkowski_scalar_2014}%
  \BibitemOpen
  \bibfield  {author} {\bibinfo {author} {\bibfnamefont {R.}~\bibnamefont
  {Wittkowski}}, \bibinfo {author} {\bibfnamefont {A.}~\bibnamefont
  {Tiribocchi}}, \bibinfo {author} {\bibfnamefont {J.}~\bibnamefont
  {Stenhammar}}, \bibinfo {author} {\bibfnamefont {R.~J.}\ \bibnamefont
  {Allen}}, \bibinfo {author} {\bibfnamefont {D.}~\bibnamefont {Marenduzzo}}, \
  and\ \bibinfo {author} {\bibfnamefont {M.~E.}\ \bibnamefont {Cates}},\
  }\bibfield  {title} {\enquote {\bibinfo {title} {{Scalar phi4 field theory
  for active-particle phase separation}},}\ }\href {\doibase
  https://doi.org/10.1038/ncomms5351} {\bibfield  {journal} {\bibinfo
  {journal} {Nat. Commun.}\ }\textbf {\bibinfo {volume} {5}},\ \bibinfo {pages}
  {145702} (\bibinfo {year} {2014})}\BibitemShut {NoStop}%
\bibitem [{\citenamefont {Solon}\ \emph {et~al.}(2015)\citenamefont {Solon},
  \citenamefont {Stenhammar}, \citenamefont {Wittkowski}, \citenamefont
  {M.~Kardar}, \citenamefont {Cates},\ and\ \citenamefont
  {Tailleur}}]{solon_pressure_2015}%
  \BibitemOpen
  \bibfield  {author} {\bibinfo {author} {\bibfnamefont {A.~P.}\ \bibnamefont
  {Solon}}, \bibinfo {author} {\bibfnamefont {J.}~\bibnamefont {Stenhammar}},
  \bibinfo {author} {\bibfnamefont {R.}~\bibnamefont {Wittkowski}}, \bibinfo
  {author} {\bibfnamefont {Y.~Kafri}\ \bibnamefont {M.~Kardar}}, \bibinfo
  {author} {\bibfnamefont {M.~E.}\ \bibnamefont {Cates}}, \ and\ \bibinfo
  {author} {\bibfnamefont {J.}~\bibnamefont {Tailleur}},\ }\bibfield  {title}
  {\enquote {\bibinfo {title} {Pressure and {Phase} {Equilibria} in
  {Interacting} {Active} {Brownian} {Spheres}},}\ }\href {\doibase
  10.1103/PhysRevLett.114.198301} {\bibfield  {journal} {\bibinfo  {journal}
  {Phys. Rev. Lett.}\ }\textbf {\bibinfo {volume} {114}},\ \bibinfo {pages}
  {198301} (\bibinfo {year} {2015})}\BibitemShut {NoStop}%
\bibitem [{\citenamefont {Solon}\ \emph {et~al.}(2018)\citenamefont {Solon},
  \citenamefont {Stenhammar}, \citenamefont {Cates}, \citenamefont {Kafri},\
  and\ \citenamefont {Tailleur}}]{solon_generalized_2018}%
  \BibitemOpen
  \bibfield  {author} {\bibinfo {author} {\bibfnamefont {A.~P.}\ \bibnamefont
  {Solon}}, \bibinfo {author} {\bibfnamefont {J.}~\bibnamefont {Stenhammar}},
  \bibinfo {author} {\bibfnamefont {M.~E.}\ \bibnamefont {Cates}}, \bibinfo
  {author} {\bibfnamefont {Y.}~\bibnamefont {Kafri}}, \ and\ \bibinfo {author}
  {\bibfnamefont {J.}~\bibnamefont {Tailleur}},\ }\bibfield  {title} {\enquote
  {\bibinfo {title} {Generalized thermodynamics of phase equilibria in scalar
  active matter},}\ }\href {\doibase 10.1103/PhysRevE.97.020602} {\bibfield
  {journal} {\bibinfo  {journal} {Phys. Rev. E}\ }\textbf {\bibinfo {volume}
  {97}},\ \bibinfo {pages} {020602} (\bibinfo {year} {2018})}\BibitemShut
  {NoStop}%
\bibitem [{\citenamefont {Tjhung}\ \emph {et~al.}(2018)\citenamefont {Tjhung},
  \citenamefont {Nardini},\ and\ \citenamefont {Cates}}]{tjhung_cluster_2018}%
  \BibitemOpen
  \bibfield  {author} {\bibinfo {author} {\bibfnamefont {E.}~\bibnamefont
  {Tjhung}}, \bibinfo {author} {\bibfnamefont {C.}~\bibnamefont {Nardini}}, \
  and\ \bibinfo {author} {\bibfnamefont {M.~E.}\ \bibnamefont {Cates}},\
  }\bibfield  {title} {\enquote {\bibinfo {title} {Cluster phases and bubbly
  phase separation in active fluids: Reversal of the ostwald process},}\ }\href
  {\doibase 10.1103/PhysRevX.8.031080} {\bibfield  {journal} {\bibinfo
  {journal} {Phys. Rev. X}\ }\textbf {\bibinfo {volume} {8}},\ \bibinfo {pages}
  {031080} (\bibinfo {year} {2018})}\BibitemShut {NoStop}%
\bibitem [{\citenamefont {Fily}\ and\ \citenamefont
  {Marchetti}(2012)}]{fily_athermal_2012}%
  \BibitemOpen
  \bibfield  {author} {\bibinfo {author} {\bibfnamefont {Y.}~\bibnamefont
  {Fily}}\ and\ \bibinfo {author} {\bibfnamefont {M.~C.}\ \bibnamefont
  {Marchetti}},\ }\bibfield  {title} {\enquote {\bibinfo {title} {Athermal
  phase separation of self-propelled particles with no alignment},}\ }\href
  {\doibase 10.1103/physrevlett.108.235702} {\bibfield  {journal} {\bibinfo
  {journal} {Phys. Rev. Lett.}\ }\textbf {\bibinfo {volume} {108}},\ \bibinfo
  {pages} {235702} (\bibinfo {year} {2012})}\BibitemShut {NoStop}%
\bibitem [{\citenamefont {Redner}\ \emph {et~al.}(2013)\citenamefont {Redner},
  \citenamefont {Hagan},\ and\ \citenamefont
  {Baskaran}}]{redner_structure_2013}%
  \BibitemOpen
  \bibfield  {author} {\bibinfo {author} {\bibfnamefont {G.~S.}\ \bibnamefont
  {Redner}}, \bibinfo {author} {\bibfnamefont {M.~F.}\ \bibnamefont {Hagan}}, \
  and\ \bibinfo {author} {\bibfnamefont {A.}~\bibnamefont {Baskaran}},\
  }\bibfield  {title} {\enquote {\bibinfo {title} {Structure and dynamics of a
  phase-separating active colloidal fluid},}\ }\href {\doibase
  10.1103/physrevlett.110.055701} {\bibfield  {journal} {\bibinfo  {journal}
  {Phys. Rev. Lett.}\ }\textbf {\bibinfo {volume} {110}},\ \bibinfo {pages}
  {055701} (\bibinfo {year} {2013})}\BibitemShut {NoStop}%
\bibitem [{\citenamefont {Speck}\ \emph {et~al.}(2014)\citenamefont {Speck},
  \citenamefont {Bialk{\'e}}, \citenamefont {Menzel},\ and\ \citenamefont
  {L{\"o}wen}}]{speck_effective_2014}%
  \BibitemOpen
  \bibfield  {author} {\bibinfo {author} {\bibfnamefont {T.}~\bibnamefont
  {Speck}}, \bibinfo {author} {\bibfnamefont {J.}~\bibnamefont {Bialk{\'e}}},
  \bibinfo {author} {\bibfnamefont {A.~M.}\ \bibnamefont {Menzel}}, \ and\
  \bibinfo {author} {\bibfnamefont {H.}~\bibnamefont {L{\"o}wen}},\ }\bibfield
  {title} {\enquote {\bibinfo {title} {Effective {Cahn}-{Hilliard} equation for
  the phase separation of active {Brownian} particles},}\ }\href {\doibase
  10.1103/physrevlett.112.218304} {\bibfield  {journal} {\bibinfo  {journal}
  {Phys. Rev. Lett.}\ }\textbf {\bibinfo {volume} {112}},\ \bibinfo {pages}
  {218304} (\bibinfo {year} {2014})}\BibitemShut {NoStop}%
\bibitem [{\citenamefont {Stenhammar}\ \emph {et~al.}(2014)\citenamefont
  {Stenhammar}, \citenamefont {Marenduzzo}, \citenamefont {Allen},\ and\
  \citenamefont {Cates}}]{stenhammar_phase_2014}%
  \BibitemOpen
  \bibfield  {author} {\bibinfo {author} {\bibfnamefont {J.}~\bibnamefont
  {Stenhammar}}, \bibinfo {author} {\bibfnamefont {D.}~\bibnamefont
  {Marenduzzo}}, \bibinfo {author} {\bibfnamefont {R.~J.}\ \bibnamefont
  {Allen}}, \ and\ \bibinfo {author} {\bibfnamefont {M.~E.}\ \bibnamefont
  {Cates}},\ }\bibfield  {title} {\enquote {\bibinfo {title} {Phase behaviour
  of active {Brownian} particles: the role of dimensionality},}\ }\href
  {\doibase 10.1039/c3sm52813h} {\bibfield  {journal} {\bibinfo  {journal}
  {Soft Matter}\ }\textbf {\bibinfo {volume} {10}},\ \bibinfo {pages}
  {1489--1499} (\bibinfo {year} {2014})}\BibitemShut {NoStop}%
\bibitem [{\citenamefont {Digregorio}\ \emph {et~al.}(2018)\citenamefont
  {Digregorio}, \citenamefont {Levis}, \citenamefont {Suma}, \citenamefont
  {Cugliandolo}, \citenamefont {Gonnella},\ and\ \citenamefont
  {Pagonabarraga}}]{digregorio_full_2018}%
  \BibitemOpen
  \bibfield  {author} {\bibinfo {author} {\bibfnamefont {P.}~\bibnamefont
  {Digregorio}}, \bibinfo {author} {\bibfnamefont {D.}~\bibnamefont {Levis}},
  \bibinfo {author} {\bibfnamefont {A.}~\bibnamefont {Suma}}, \bibinfo {author}
  {\bibfnamefont {L.~F.}\ \bibnamefont {Cugliandolo}}, \bibinfo {author}
  {\bibfnamefont {G.}~\bibnamefont {Gonnella}}, \ and\ \bibinfo {author}
  {\bibfnamefont {I.}~\bibnamefont {Pagonabarraga}},\ }\bibfield  {title}
  {\enquote {\bibinfo {title} {Full phase diagram of active {Brownian} disks:
  {From} melting to motility-induced phase separation},}\ }\href {\doibase
  10.1103/physrevlett.121.098003} {\bibfield  {journal} {\bibinfo  {journal}
  {Phys. Rev. Lett.}\ }\textbf {\bibinfo {volume} {121}},\ \bibinfo {pages}
  {098003} (\bibinfo {year} {2018})}\BibitemShut {NoStop}%
\bibitem [{\citenamefont {Klamser}\ \emph {et~al.}(2018)\citenamefont
  {Klamser}, \citenamefont {Kapfer},\ and\ \citenamefont
  {Krauth}}]{klamser_thermodynamic_2018}%
  \BibitemOpen
  \bibfield  {author} {\bibinfo {author} {\bibfnamefont {J.~U.}\ \bibnamefont
  {Klamser}}, \bibinfo {author} {\bibfnamefont {S.~C.}\ \bibnamefont {Kapfer}},
  \ and\ \bibinfo {author} {\bibfnamefont {W.}~\bibnamefont {Krauth}},\
  }\bibfield  {title} {\enquote {\bibinfo {title} {Thermodynamic phases in
  two-dimensional active matter},}\ }\href {\doibase
  10.1038/s41467-018-07491-5} {\bibfield  {journal} {\bibinfo  {journal} {Nat.
  Commun.}\ }\textbf {\bibinfo {volume} {9}},\ \bibinfo {pages} {5045}
  (\bibinfo {year} {2018})}\BibitemShut {NoStop}%
\bibitem [{\citenamefont {Mandal}\ \emph {et~al.}(2019)\citenamefont {Mandal},
  \citenamefont {Liebchen},\ and\ \citenamefont
  {L\"owen}}]{mandal_motility_2019}%
  \BibitemOpen
  \bibfield  {author} {\bibinfo {author} {\bibfnamefont {S.}~\bibnamefont
  {Mandal}}, \bibinfo {author} {\bibfnamefont {B.}~\bibnamefont {Liebchen}}, \
  and\ \bibinfo {author} {\bibfnamefont {H.}~\bibnamefont {L\"owen}},\
  }\bibfield  {title} {\enquote {\bibinfo {title} {Motility-induced temperature
  difference in coexisting phases},}\ }\href {\doibase
  10.1103/PhysRevLett.123.228001} {\bibfield  {journal} {\bibinfo  {journal}
  {Phys. Rev. Lett.}\ }\textbf {\bibinfo {volume} {123}},\ \bibinfo {pages}
  {228001} (\bibinfo {year} {2019})}\BibitemShut {NoStop}%
\bibitem [{\citenamefont {Caprini}\ \emph {et~al.}(2020)\citenamefont
  {Caprini}, \citenamefont {Marconi},\ and\ \citenamefont
  {Puglisi}}]{caprini_spontaneous_2020}%
  \BibitemOpen
  \bibfield  {author} {\bibinfo {author} {\bibfnamefont {L.}~\bibnamefont
  {Caprini}}, \bibinfo {author} {\bibfnamefont {U.~Marini~Bettolo}\
  \bibnamefont {Marconi}}, \ and\ \bibinfo {author} {\bibfnamefont
  {A.}~\bibnamefont {Puglisi}},\ }\bibfield  {title} {\enquote {\bibinfo
  {title} {Spontaneous velocity alignment in motility-induced phase
  separation},}\ }\href {\doibase 10.1103/PhysRevLett.124.078001} {\bibfield
  {journal} {\bibinfo  {journal} {Phys. Rev. Lett.}\ }\textbf {\bibinfo
  {volume} {124}},\ \bibinfo {pages} {078001} (\bibinfo {year}
  {2020})}\BibitemShut {NoStop}%
\bibitem [{\citenamefont {Liu}\ \emph {et~al.}(2019)\citenamefont {Liu},
  \citenamefont {Patch}, \citenamefont {Bahar}, \citenamefont {Yllanes},
  \citenamefont {Welch}, \citenamefont {Marchetti}, \citenamefont
  {Thutupalli},\ and\ \citenamefont {Shaevitz}}]{liu_self-driven_2019}%
  \BibitemOpen
  \bibfield  {author} {\bibinfo {author} {\bibfnamefont {G.}~\bibnamefont
  {Liu}}, \bibinfo {author} {\bibfnamefont {A.}~\bibnamefont {Patch}}, \bibinfo
  {author} {\bibfnamefont {F.}~\bibnamefont {Bahar}}, \bibinfo {author}
  {\bibfnamefont {D.}~\bibnamefont {Yllanes}}, \bibinfo {author} {\bibfnamefont
  {R.}~\bibnamefont {Welch}}, \bibinfo {author} {\bibfnamefont {M.~C.}\
  \bibnamefont {Marchetti}}, \bibinfo {author} {\bibfnamefont {S.}~\bibnamefont
  {Thutupalli}}, \ and\ \bibinfo {author} {\bibfnamefont {J.~W.}\ \bibnamefont
  {Shaevitz}},\ }\bibfield  {title} {\enquote {\bibinfo {title} {Self-{Driven}
  {Phase} {Transitions} {Drive} {Myxococcus} xanthus {Fruiting} {Body}
  {Formation}},}\ }\href {\doibase 10.1103/physrevlett.122.248102} {\bibfield
  {journal} {\bibinfo  {journal} {Phys. Rev. Lett.}\ }\textbf {\bibinfo
  {volume} {122}},\ \bibinfo {pages} {248102} (\bibinfo {year}
  {2019})}\BibitemShut {NoStop}%
\bibitem [{\citenamefont {Buttinoni}\ \emph {et~al.}(2013)\citenamefont
  {Buttinoni}, \citenamefont {Bialk{\'e}}, \citenamefont {K{\"u}mmel},
  \citenamefont {L{\"o}wen}, \citenamefont {Bechinger},\ and\ \citenamefont
  {Speck}}]{buttinoni_dynamical_2013}%
  \BibitemOpen
  \bibfield  {author} {\bibinfo {author} {\bibfnamefont {I.}~\bibnamefont
  {Buttinoni}}, \bibinfo {author} {\bibfnamefont {J.}~\bibnamefont
  {Bialk{\'e}}}, \bibinfo {author} {\bibfnamefont {F.}~\bibnamefont
  {K{\"u}mmel}}, \bibinfo {author} {\bibfnamefont {H.}~\bibnamefont
  {L{\"o}wen}}, \bibinfo {author} {\bibfnamefont {C.}~\bibnamefont
  {Bechinger}}, \ and\ \bibinfo {author} {\bibfnamefont {T.}~\bibnamefont
  {Speck}},\ }\bibfield  {title} {\enquote {\bibinfo {title} {Dynamical
  clustering and phase separation in suspensions of self-propelled colloidal
  particles},}\ }\href {\doibase 10.1103/physrevlett.110.238301} {\bibfield
  {journal} {\bibinfo  {journal} {Phys. Rev. Lett.}\ }\textbf {\bibinfo
  {volume} {110}},\ \bibinfo {pages} {238301} (\bibinfo {year}
  {2013})}\BibitemShut {NoStop}%
\bibitem [{\citenamefont {Geyer}\ \emph {et~al.}(2019)\citenamefont {Geyer},
  \citenamefont {Martin}, \citenamefont {Tailleur},\ and\ \citenamefont
  {Bartolo}}]{geyer_freezing_2019}%
  \BibitemOpen
  \bibfield  {author} {\bibinfo {author} {\bibfnamefont {D.}~\bibnamefont
  {Geyer}}, \bibinfo {author} {\bibfnamefont {D.}~\bibnamefont {Martin}},
  \bibinfo {author} {\bibfnamefont {J.}~\bibnamefont {Tailleur}}, \ and\
  \bibinfo {author} {\bibfnamefont {D.}~\bibnamefont {Bartolo}},\ }\bibfield
  {title} {\enquote {\bibinfo {title} {Freezing a flock: {Motility}-induced
  phase separation in polar active liquids},}\ }\href {\doibase
  10.1103/physrevx.9.031043} {\bibfield  {journal} {\bibinfo  {journal} {Phys.
  Rev. X}\ }\textbf {\bibinfo {volume} {9}},\ \bibinfo {pages} {031043}
  (\bibinfo {year} {2019})}\BibitemShut {NoStop}%
\bibitem [{\citenamefont {van~der Linden}\ \emph {et~al.}(2019)\citenamefont
  {van~der Linden}, \citenamefont {Alexander}, \citenamefont {Aarts.},\ and\
  \citenamefont {Dauchot}}]{vanderlinden_interrupted_2019}%
  \BibitemOpen
  \bibfield  {author} {\bibinfo {author} {\bibfnamefont {M.~N.}\ \bibnamefont
  {van~der Linden}}, \bibinfo {author} {\bibfnamefont {L.~C.}\ \bibnamefont
  {Alexander}}, \bibinfo {author} {\bibfnamefont {D.~G.}\ \bibnamefont
  {Aarts.}}, \ and\ \bibinfo {author} {\bibfnamefont {O.}~\bibnamefont
  {Dauchot}},\ }\bibfield  {title} {\enquote {\bibinfo {title} {Interrupted
  motility induced phase separation in aligning active colloids},}\ }\href
  {\doibase 10.1103/PhysRevLett.123.098001} {\bibfield  {journal} {\bibinfo
  {journal} {Phys. Rev. Lett.}\ }\textbf {\bibinfo {volume} {123}},\ \bibinfo
  {pages} {098001} (\bibinfo {year} {2019})}\BibitemShut {NoStop}%
\bibitem [{\citenamefont {{Solon}}\ \emph {et~al.}(2018)\citenamefont
  {{Solon}}, \citenamefont {{Stenhammar}}, \citenamefont {{Cates}},
  \citenamefont {{Kafri}},\ and\ \citenamefont
  {{Tailleur}}}]{solon_generalized_2018-1}%
  \BibitemOpen
  \bibfield  {author} {\bibinfo {author} {\bibfnamefont {A.~P.}\ \bibnamefont
  {{Solon}}}, \bibinfo {author} {\bibfnamefont {J.}~\bibnamefont
  {{Stenhammar}}}, \bibinfo {author} {\bibfnamefont {M.~E.}\ \bibnamefont
  {{Cates}}}, \bibinfo {author} {\bibfnamefont {Y.}~\bibnamefont {{Kafri}}}, \
  and\ \bibinfo {author} {\bibfnamefont {J.}~\bibnamefont {{Tailleur}}},\
  }\bibfield  {title} {\enquote {\bibinfo {title} {{Generalized thermodynamics
  of motility-induced phase separation: phase equilibria, Laplace pressure, and
  change of ensembles}},}\ }\href {\doibase 10.1088/1367-2630/aaccdd}
  {\bibfield  {journal} {\bibinfo  {journal} {New J. Phys.}\ }\textbf {\bibinfo
  {volume} {20}},\ \bibinfo {eid} {075001} (\bibinfo {year}
  {2018})}\BibitemShut {NoStop}%
\bibitem [{\citenamefont {{Solon}}\ \emph {et~al.}(2015)\citenamefont
  {{Solon}}, \citenamefont {{Fily}}, \citenamefont {{Baskaran}}, \citenamefont
  {{Cates}}, \citenamefont {{Kafri}}, \citenamefont {{Kardar}},\ and\
  \citenamefont {{Tailleur}}}]{solon_pressure_2015-2}%
  \BibitemOpen
  \bibfield  {author} {\bibinfo {author} {\bibfnamefont {A.~P.}\ \bibnamefont
  {{Solon}}}, \bibinfo {author} {\bibfnamefont {Y.}~\bibnamefont {{Fily}}},
  \bibinfo {author} {\bibfnamefont {A.}~\bibnamefont {{Baskaran}}}, \bibinfo
  {author} {\bibfnamefont {M.~E.}\ \bibnamefont {{Cates}}}, \bibinfo {author}
  {\bibfnamefont {Y.}~\bibnamefont {{Kafri}}}, \bibinfo {author} {\bibfnamefont
  {M.}~\bibnamefont {{Kardar}}}, \ and\ \bibinfo {author} {\bibfnamefont
  {J.}~\bibnamefont {{Tailleur}}},\ }\bibfield  {title} {\enquote {\bibinfo
  {title} {{Pressure is not a state function for generic active fluids}},}\
  }\href {\doibase 10.1038/nphys3377} {\bibfield  {journal} {\bibinfo
  {journal} {Nat. Phys.}\ }\textbf {\bibinfo {volume} {11}},\ \bibinfo {pages}
  {673--678} (\bibinfo {year} {2015})}\BibitemShut {NoStop}%
\bibitem [{\citenamefont {Bialk\'e}\ \emph {et~al.}(2015)\citenamefont
  {Bialk\'e}, \citenamefont {Siebert}, \citenamefont {L\"owen},\ and\
  \citenamefont {Speck}}]{bialke_negative_2015}%
  \BibitemOpen
  \bibfield  {author} {\bibinfo {author} {\bibfnamefont {J.}~\bibnamefont
  {Bialk\'e}}, \bibinfo {author} {\bibfnamefont {J.~T.}\ \bibnamefont
  {Siebert}}, \bibinfo {author} {\bibfnamefont {H.}~\bibnamefont {L\"owen}}, \
  and\ \bibinfo {author} {\bibfnamefont {T.}~\bibnamefont {Speck}},\ }\bibfield
   {title} {\enquote {\bibinfo {title} {Negative interfacial tension in
  phase-separated active brownian particles},}\ }\href {\doibase
  10.1103/PhysRevLett.115.098301} {\bibfield  {journal} {\bibinfo  {journal}
  {Phys. Rev. Lett.}\ }\textbf {\bibinfo {volume} {115}},\ \bibinfo {pages}
  {098301} (\bibinfo {year} {2015})}\BibitemShut {NoStop}%
\bibitem [{\citenamefont {Patch}\ \emph {et~al.}(2018)\citenamefont {Patch},
  \citenamefont {Sussman}, \citenamefont {Yllanes},\ and\ \citenamefont
  {Marchetti}}]{patch_curvature-dependent_2018}%
  \BibitemOpen
  \bibfield  {author} {\bibinfo {author} {\bibfnamefont {A.}~\bibnamefont
  {Patch}}, \bibinfo {author} {\bibfnamefont {D.~M.}\ \bibnamefont {Sussman}},
  \bibinfo {author} {\bibfnamefont {D.}~\bibnamefont {Yllanes}}, \ and\
  \bibinfo {author} {\bibfnamefont {M.~C.}\ \bibnamefont {Marchetti}},\
  }\bibfield  {title} {\enquote {\bibinfo {title} {Curvature-dependent tension
  and tangential flows at the interface of motility-induced phases},}\ }\href
  {\doibase 10.1039/c8sm00899j} {\bibfield  {journal} {\bibinfo  {journal}
  {Soft matter}\ }\textbf {\bibinfo {volume} {14}},\ \bibinfo {pages}
  {7435--7445} (\bibinfo {year} {2018})}\BibitemShut {NoStop}%
\bibitem [{Note1()}]{Note1}%
  \BibitemOpen
  \bibinfo {note} {Two recent papers either contest~\cite
  {hermann_non-negative_2019} or regard as physically irrelevant~\cite
  {omar_microscopic_2020} that the surface tension is measured to be negative
  in numerical simulations.}\BibitemShut {Stop}%
\bibitem [{\citenamefont {Caporusso}\ \emph {et~al.}(2020)\citenamefont
  {Caporusso}, \citenamefont {Digregorio}, \citenamefont {Levis}, \citenamefont
  {Cugliandolo},\ and\ \citenamefont {Gonnella}}]{caporusso2020micro}%
  \BibitemOpen
  \bibfield  {author} {\bibinfo {author} {\bibfnamefont {C.~B.}\ \bibnamefont
  {Caporusso}}, \bibinfo {author} {\bibfnamefont {P.}~\bibnamefont
  {Digregorio}}, \bibinfo {author} {\bibfnamefont {D.}~\bibnamefont {Levis}},
  \bibinfo {author} {\bibfnamefont {L.~F.}\ \bibnamefont {Cugliandolo}}, \ and\
  \bibinfo {author} {\bibfnamefont {G.}~\bibnamefont {Gonnella}},\ }\href@noop
  {} {\enquote {\bibinfo {title} {Micro and macro motility-induced phase
  separation in a two-dimensional active brownian particle system},}\ }
  (\bibinfo {year} {2020}),\ \Eprint {http://arxiv.org/abs/2005.06893}
  {arXiv:2005.06893 [cond-mat.stat-mech]} \BibitemShut {NoStop}%
\bibitem [{SUP()}]{SUPP}%
  \BibitemOpen
  \href@noop {} {}\bibinfo {note} {See supplementary information
  online.}\BibitemShut {Stop}%
\bibitem [{Note2()}]{Note2}%
  \BibitemOpen
  \bibinfo {note} {This may explain why previous works,all performed with
  isotropic mobility, did not pay much attention to bubbles that remained rare
  and small at the sizes studied.}\BibitemShut {Stop}%
\bibitem [{Note3()}]{Note3}%
  \BibitemOpen
  \bibinfo {note} {This is illustrated in Fig.~\ref {fig3}(a) for the other
  model studied here.}\BibitemShut {Stop}%
\bibitem [{\citenamefont {Thompson}\ \emph {et~al.}(2011)\citenamefont
  {Thompson}, \citenamefont {Tailleur}, \citenamefont {Cates},\ and\
  \citenamefont {Blythe}}]{thompson_lattice_2011}%
  \BibitemOpen
  \bibfield  {author} {\bibinfo {author} {\bibfnamefont {A.~G.}\ \bibnamefont
  {Thompson}}, \bibinfo {author} {\bibfnamefont {J.}~\bibnamefont {Tailleur}},
  \bibinfo {author} {\bibfnamefont {M.~E.}\ \bibnamefont {Cates}}, \ and\
  \bibinfo {author} {\bibfnamefont {R.~A.}\ \bibnamefont {Blythe}},\ }\bibfield
   {title} {\enquote {\bibinfo {title} {Lattice models of nonequilibrium
  bacterial dynamics},}\ }\href {\doibase 10.1088/1742-5468/2011/02/p02029}
  {\bibfield  {journal} {\bibinfo  {journal} {J. Stat. Mech.: Theory Exp}\
  }\textbf {\bibinfo {volume} {2011}},\ \bibinfo {pages} {P02029} (\bibinfo
  {year} {2011})}\BibitemShut {NoStop}%
\bibitem [{\citenamefont {Soto}\ and\ \citenamefont
  {Golestanian}(2014)}]{soto2014self}%
  \BibitemOpen
  \bibfield  {author} {\bibinfo {author} {\bibfnamefont {R.}~\bibnamefont
  {Soto}}\ and\ \bibinfo {author} {\bibfnamefont {R.}~\bibnamefont
  {Golestanian}},\ }\bibfield  {title} {\enquote {\bibinfo {title}
  {Self-assembly of catalytically active colloidal molecules: Tailoring
  activity through surface chemistry},}\ }\href {\doibase
  10.1103/physrevlett.112.068301} {\bibfield  {journal} {\bibinfo  {journal}
  {Phys. Rev. Lett.}\ }\textbf {\bibinfo {volume} {112}},\ \bibinfo {pages}
  {068301} (\bibinfo {year} {2014})}\BibitemShut {NoStop}%
\bibitem [{\citenamefont {Sep{\'u}lveda}\ and\ \citenamefont
  {Soto}(2017)}]{sepulveda_wetting_2017}%
  \BibitemOpen
  \bibfield  {author} {\bibinfo {author} {\bibfnamefont {N.}~\bibnamefont
  {Sep{\'u}lveda}}\ and\ \bibinfo {author} {\bibfnamefont {R.}~\bibnamefont
  {Soto}},\ }\bibfield  {title} {\enquote {\bibinfo {title} {Wetting
  transitions displayed by persistent active particles},}\ }\href {\doibase
  10.1103/physrevlett.119.078001} {\bibfield  {journal} {\bibinfo  {journal}
  {Phys. Rev. Lett.}\ }\textbf {\bibinfo {volume} {119}},\ \bibinfo {pages}
  {078001} (\bibinfo {year} {2017})}\BibitemShut {NoStop}%
\bibitem [{\citenamefont {Whitelam}\ \emph {et~al.}(2018)\citenamefont
  {Whitelam}, \citenamefont {Klymko},\ and\ \citenamefont
  {Mandal}}]{whitelam_phase_2018}%
  \BibitemOpen
  \bibfield  {author} {\bibinfo {author} {\bibfnamefont {S.}~\bibnamefont
  {Whitelam}}, \bibinfo {author} {\bibfnamefont {K.}~\bibnamefont {Klymko}}, \
  and\ \bibinfo {author} {\bibfnamefont {D.}~\bibnamefont {Mandal}},\
  }\bibfield  {title} {\enquote {\bibinfo {title} {Phase separation and large
  deviations of lattice active matter},}\ }\href {\doibase 10.1063/1.5023403}
  {\bibfield  {journal} {\bibinfo  {journal} {J. Chem. Phys}\ }\textbf
  {\bibinfo {volume} {148}},\ \bibinfo {pages} {154902} (\bibinfo {year}
  {2018})}\BibitemShut {NoStop}%
\bibitem [{\citenamefont {Partridge}\ and\ \citenamefont
  {Lee}(2019)}]{partridge_critical_2019}%
  \BibitemOpen
  \bibfield  {author} {\bibinfo {author} {\bibfnamefont {B.}~\bibnamefont
  {Partridge}}\ and\ \bibinfo {author} {\bibfnamefont {C.~F.}\ \bibnamefont
  {Lee}},\ }\bibfield  {title} {\enquote {\bibinfo {title} {Critical
  motility-induced phase separation belongs to the {Ising} universality
  class},}\ }\href {\doibase 10.1103/physrevlett.123.068002} {\bibfield
  {journal} {\bibinfo  {journal} {Phys. Rev. Lett.}\ }\textbf {\bibinfo
  {volume} {123}},\ \bibinfo {pages} {068002} (\bibinfo {year}
  {2019})}\BibitemShut {NoStop}%
\bibitem [{\citenamefont {Pruessner}(2012)}]{Pruessner_SOC_2012}%
  \BibitemOpen
  \bibfield  {author} {\bibinfo {author} {\bibfnamefont {G.}~\bibnamefont
  {Pruessner}},\ }\href@noop {} {\emph {\bibinfo {title} {{Self-Organised
  Criticality}}}}\ (\bibinfo  {publisher} {Cambridge University Press},\
  \bibinfo {year} {2012})\BibitemShut {NoStop}%
\bibitem [{\citenamefont {Buend{\'\i}a}\ \emph {et~al.}(2020)\citenamefont
  {Buend{\'\i}a}, \citenamefont {di~Santo}, \citenamefont {Bonachela},\ and\
  \citenamefont {Mu{\~n}oz}}]{buenda2020feedback}%
  \BibitemOpen
  \bibfield  {author} {\bibinfo {author} {\bibfnamefont {V.}~\bibnamefont
  {Buend{\'\i}a}}, \bibinfo {author} {\bibfnamefont {S.}~\bibnamefont
  {di~Santo}}, \bibinfo {author} {\bibfnamefont {J.~A.}\ \bibnamefont
  {Bonachela}}, \ and\ \bibinfo {author} {\bibfnamefont {M.~A.}\ \bibnamefont
  {Mu{\~n}oz}},\ }\href@noop {} {\enquote {\bibinfo {title} {{Feedback
  mechanisms for self-organization to the edge of a phase transition}},}\ }
  (\bibinfo {year} {2020}),\ \Eprint {http://arxiv.org/abs/2006.03020}
  {arXiv:2006.03020 [cond-mat.stat-mech]} \BibitemShut {NoStop}%
\bibitem [{Note4()}]{Note4}%
  \BibitemOpen
  \bibinfo {note} {For a similar approach in a different context, see \cite
  {ranft_aggregation_2017}.}\BibitemShut {Stop}%
\bibitem [{\citenamefont {Bray}(2002)}]{bray2002theory}%
  \BibitemOpen
  \bibfield  {author} {\bibinfo {author} {\bibfnamefont {A.~J}\ \bibnamefont
  {Bray}},\ }\bibfield  {title} {\enquote {\bibinfo {title} {Theory of
  phase-ordering kinetics},}\ }\href {\doibase 10.1080/00018730110117433}
  {\bibfield  {journal} {\bibinfo  {journal} {Advances in Physics}\ }\textbf
  {\bibinfo {volume} {51}},\ \bibinfo {pages} {481--587} (\bibinfo {year}
  {2002})}\BibitemShut {NoStop}%
\bibitem [{\citenamefont {Siebert}\ \emph {et~al.}(2018)\citenamefont
  {Siebert}, \citenamefont {Dittrich}, \citenamefont {Schmid}, \citenamefont
  {Binder}, \citenamefont {Speck},\ and\ \citenamefont
  {Virnau}}]{siebert_critical_2018}%
  \BibitemOpen
  \bibfield  {author} {\bibinfo {author} {\bibfnamefont {J.~T.}\ \bibnamefont
  {Siebert}}, \bibinfo {author} {\bibfnamefont {F.}~\bibnamefont {Dittrich}},
  \bibinfo {author} {\bibfnamefont {F.}~\bibnamefont {Schmid}}, \bibinfo
  {author} {\bibfnamefont {K.}~\bibnamefont {Binder}}, \bibinfo {author}
  {\bibfnamefont {T.}~\bibnamefont {Speck}}, \ and\ \bibinfo {author}
  {\bibfnamefont {P.}~\bibnamefont {Virnau}},\ }\bibfield  {title} {\enquote
  {\bibinfo {title} {Critical behavior of active {Brownian} particles},}\
  }\href {\doibase 10.1103/physreve.98.030601} {\bibfield  {journal} {\bibinfo
  {journal} {Phys. Rev. E}\ }\textbf {\bibinfo {volume} {98}},\ \bibinfo
  {pages} {030601} (\bibinfo {year} {2018})}\BibitemShut {NoStop}%
\bibitem [{\citenamefont {Hermann}\ \emph {et~al.}(2019)\citenamefont
  {Hermann}, \citenamefont {de~las Heras},\ and\ \citenamefont
  {Schmidt}}]{hermann_non-negative_2019}%
  \BibitemOpen
  \bibfield  {author} {\bibinfo {author} {\bibfnamefont {S.}~\bibnamefont
  {Hermann}}, \bibinfo {author} {\bibfnamefont {D.}~\bibnamefont {de~las
  Heras}}, \ and\ \bibinfo {author} {\bibfnamefont {M.}~\bibnamefont
  {Schmidt}},\ }\bibfield  {title} {\enquote {\bibinfo {title} {Non-negative
  {Interfacial} {Tension} in {Phase}-{Separated} {Active} {Brownian}
  {Particles}},}\ }\href {\doibase 10.1103/physrevlett.123.268002} {\bibfield
  {journal} {\bibinfo  {journal} {Phys. Rev. Lett.}\ }\textbf {\bibinfo
  {volume} {123}},\ \bibinfo {pages} {268002} (\bibinfo {year}
  {2019})}\BibitemShut {NoStop}%
\bibitem [{\citenamefont {Omar}\ \emph {et~al.}(2020)\citenamefont {Omar},
  \citenamefont {Wang},\ and\ \citenamefont {Brady}}]{omar_microscopic_2020}%
  \BibitemOpen
  \bibfield  {author} {\bibinfo {author} {\bibfnamefont {A.~K.}\ \bibnamefont
  {Omar}}, \bibinfo {author} {\bibfnamefont {Z.-G.}\ \bibnamefont {Wang}}, \
  and\ \bibinfo {author} {\bibfnamefont {J.~F.}\ \bibnamefont {Brady}},\
  }\bibfield  {title} {\enquote {\bibinfo {title} {Microscopic origins of the
  swim pressure and the anomalous surface tension of active matter},}\ }\href
  {\doibase 10.1103/physreve.101.012604} {\bibfield  {journal} {\bibinfo
  {journal} {Phys. Rev. E}\ }\textbf {\bibinfo {volume} {101}},\ \bibinfo
  {pages} {012604} (\bibinfo {year} {2020})}\BibitemShut {NoStop}%
\bibitem [{\citenamefont {{Ranft}}\ \emph {et~al.}(2017)\citenamefont
  {{Ranft}}, \citenamefont {{Almeida}}, \citenamefont {{Rodriguez}},
  \citenamefont {{Triller}},\ and\ \citenamefont
  {{Hakim}}}]{ranft_aggregation_2017}%
  \BibitemOpen
  \bibfield  {author} {\bibinfo {author} {\bibfnamefont {J.}~\bibnamefont
  {{Ranft}}}, \bibinfo {author} {\bibfnamefont {L.~G.}\ \bibnamefont
  {{Almeida}}}, \bibinfo {author} {\bibfnamefont {P.~C.}\ \bibnamefont
  {{Rodriguez}}}, \bibinfo {author} {\bibfnamefont {A.}~\bibnamefont
  {{Triller}}}, \ and\ \bibinfo {author} {\bibfnamefont {V.}~\bibnamefont
  {{Hakim}}},\ }\bibfield  {title} {\enquote {\bibinfo {title} {{An
  aggregation-removal model for the formation and size determination of
  post-synaptic scaffold domains}},}\ }\href {\doibase
  10.1371/journal.pcbi.1005516} {\bibfield  {journal} {\bibinfo  {journal}
  {PLoS Computational Biology}\ }\textbf {\bibinfo {volume} {13}},\ \bibinfo
  {pages} {e1005516} (\bibinfo {year} {2017})}\BibitemShut {NoStop}%
\end{thebibliography}%

\end{document}